\documentclass{emulateapj-rtx4}

\usepackage{epsf}
\usepackage{graphicx}
\usepackage{lscape}
\usepackage{apjfonts}

\newcommand{\bjdtdb}{\ensuremath{\rm {BJD_{TDB}}}}
\newcommand{\feh}{\ensuremath{\left[{\rm Fe}/{\rm H}\right]}}
\newcommand{\teff}{\ensuremath{T_{\rm eff}}}
\newcommand{\ecosw}{\ensuremath{e\cos{\omega_*}}}
\newcommand{\esinw}{\ensuremath{e\sin{\omega_*}}}
\newcommand{\msun}{\ensuremath{\,{\rm M}_\Sun}}
\newcommand{\rsun}{\ensuremath{\,{\rm R}_\Sun}}
\newcommand{\lsun}{\ensuremath{\,{\rm L}_\Sun}}
\newcommand{\mj}{\ensuremath{\,{\rm M}_{\rm J}}}
\newcommand{\rj}{\ensuremath{\,{\rm R}_{\rm J}}}
\newcommand{\fave}{\langle F \rangle}
\newcommand{\fluxcgs}{10$^9$ erg s$^{-1}$ cm$^{-2}$}

\begin{document}
\title{KELT-3b: A Hot Jupiter Transiting a $V=9.8$ Late-F Star}
\author{Joshua Pepper\altaffilmark{1,2},
Robert J.\ Siverd\altaffilmark{2},
Thomas G.\ Beatty\altaffilmark{3},
B.\ Scott Gaudi\altaffilmark{3},
Keivan G.\ Stassun\altaffilmark{2,4},
Jason Eastman\altaffilmark{5,6},
Karen Collins\altaffilmark{7},
David W. Latham\altaffilmark{8},
Allyson Bieryla\altaffilmark{8},            
Lars A. Buchhave\altaffilmark{9,10},
Eric L. N. Jensen\altaffilmark{11},
Mark Manner\altaffilmark{12},
Kaloyan Penev\altaffilmark{13},
Justin R. Crepp\altaffilmark{14},
Phillip A. Cargile\altaffilmark{2},
Saurav Dhital\altaffilmark{2,15},
Michael L. Calkins\altaffilmark{8},
Gilbert A. Esquerdo\altaffilmark{8},
Perry Berlind\altaffilmark{8},
Benjamin J. Fulton\altaffilmark{5,16},
Rachel Street\altaffilmark{5},
Bo Ma\altaffilmark{17},
Jian Ge\altaffilmark{17},
Ji Wang\altaffilmark{18},
Qingqing Mao\altaffilmark{2},
Alexander J. W. Richert\altaffilmark{19},
Andrew Gould\altaffilmark{3,20},
Darren L. DePoy\altaffilmark{21},
John F. Kielkopf\altaffilmark{7},
Jennifer L. Marshall\altaffilmark{21},
Richard W. Pogge\altaffilmark{3,20},
Robert P. Stefanik\altaffilmark{8},
Mark Trueblood\altaffilmark{22},
Patricia Trueblood\altaffilmark{22}
}

\altaffiltext{1}{Department of Physics, Lehigh University, Bethlehem, PA 18015, USA}
\altaffiltext{2}{Department of Physics and Astronomy, Vanderbilt University, Nashville, TN 37235, USA}
\altaffiltext{3}{Department of Astronomy, The Ohio State University, Columbus, OH 43210, USA}
\altaffiltext{4}{Department of Physics, Fisk University, Nashville, TN 37208, USA}
\altaffiltext{5}{Las Cumbres Observatory Global Telescope Network, Santa Barbara, CA 93117, USA}
\altaffiltext{6}{Department of Physics Broida Hall, University of California, Santa Barbara, CA 93106, USA}
\altaffiltext{7}{Department of Physics \& Astronomy, University of Louisville, Louisville, KY 40292, USA}
\altaffiltext{8}{Harvard-Smithsonian Center for Astrophysics, Cambridge, MA 02138, USA}
\altaffiltext{9}{Niels Bohr Institute, University of Copenhagen, 21S00 Copenhagen, Denmark}
\altaffiltext{10}{Centre for Star and Planet Formation, Natural History Museum of Denmark, University of Copenhagen, DK-1350 Copenhagen, Denmark}
\altaffiltext{11}{Department of Physics and Astronomy, Swarthmore College, Swarthmore, PA 19081, USA}
\altaffiltext{12}{Spot Observatory, Nunnelly, TN 37137, USA}
\altaffiltext{13}{Department of Astrophysical Sciences, Princeton University, Peyton Hall, Princeton, NJ 08544, USA}
\altaffiltext{14}{Department of Physics, University of Notre Dame, Notre Dame, IN 46556, USA}
\altaffiltext{15}{Department of Astronomy, Boston University, 725 Commonwealth Avenue, Boston, MA 02215, USA}
\altaffiltext{16}{Institute for Astronomy, University of Hawaii, Honolulu, HI 96822, USA}
\altaffiltext{17}{Department of Astronomy, University of Florida, Gainesville, FL 32611, USA}
\altaffiltext{18}{Department of Astronomy, Yale University, New Haven, CT 06511 USA }
\altaffiltext{19}{Department of Astronomy and Astrophysics, Pennsylvania State University, University Park, PA 16802, USA}
\altaffiltext{20}{Center for Cosmology and Astroparticle Physics, The Ohio State University, OH 43210, USA}
\altaffiltext{21}{Department of Physics \& Astronomy, Texas A\&M University, College Station, TX 77843, USA}
\altaffiltext{22}{Winer Observatory, Sonoita, AZ 85637, USA}

\begin{abstract}
We report the discovery of KELT-3b, a moderately inflated transiting hot Jupiter with a mass of $1.477_{-0.067}^{+0.066} M_J$, and radius of $1.345\pm0.072 R_J$, with an orbital period of $2.7033904\pm0.000010$ days.  The host star, KELT-3, is a $V=9.8$ late F star with $M_* = 1.278_{-0.061}^{+0.063} M_{\odot}$, $R_* = 1.472_{-0.067}^{+0.065} R_{\odot}$, $T_{\rm eff} = 6306_{-49}^{+50}$ K, log(g) = $4.209_{-0.031}^{+0.033}$, and [Fe/H] = $0.044_{-0.082}^{+0.080}$, and has a likely proper motion companion. KELT-3b is the third transiting exoplanet discovered by the KELT survey, and is orbiting one of the 20 brightest known transiting planet host stars, making it a promising candidate for detailed characterization studies.  Although we infer that KELT-3 is significantly evolved, a preliminary analysis of the stellar and orbital evolution of the system suggests that the planet has likely always received a level of incident flux above the empirically-identified threshold for radius inflation suggested by \citet{ds11}.
\end{abstract}
    
\bibliographystyle{apj}
\keywords{}

\section{Introduction\label{sec:intro}}

Transiting extrasolar planets are the best laboratories for studying the individual properties of exoplanets, providing clues about planetary formation and evolution.  Information about planetary mass, radius, atmosphere, and spin-orbit alignment can most easily be gathered from transiting exoplanets that orbit bright stars, since the greater flux from such stars enables faster, cheaper, and more precise follow-up observations.

Several ground-based transit surveys (TrES, \citet{alon04}; XO, \citet{mc06}; HATNet, \citet{bakos07}; SuperWASP, \citet{cc07a}; QES, \citet{al07}) have been conducted, some of which are still ongoing, and have produced a large number of exoplanet discoveries.  SuperWASP and HATNet have been especially productive, with each survey discovering dozens of new transiting planets.  The CoRoT \citep{bag03} and Kepler \citep{boru10} space-based missions have also made numerous exoplanet discoveries, and have expanded the parameter space probed for transiting planets, finding those with both very long periods and much smaller radii.

The Kilodegree Extremely Little Telescope (KELT) transit survey is designed to find transiting planets around bright stars.  The KELT-North telescope \citep{pepper07} has a small aperture and wide field of view ($26^{\circ} \times 26^{\circ}$) to observe the entire sky between declinations $19^{\circ}$N and $45^{\circ}$N, covering approximately 40\% of the northern sky.   The aperture, optical system, and exposure time for KELT-North are configured to obtain better than 1\% RMS photometry for stars with $8 < V < 10$.  That magnitude range represents the brightness gap between comprehensive RV surveys and most other transit surveys.  Specifically, while both SuperWASP and HATNet have been tremendously successful at discovering transiting exoplanets, the vast majority of their published discoveries (90\% and 93\%, respectively) are fainter than $V=10$.  On the other hand, RV surveys start to become incomplete for stars fainter than $V\sim8$ \citep{wright12}.

The KELT-North survey has been operating since 2006, and we have been vetting transit candidates since April 2011.  The first two KELT transit discoveries demonstrate the scientific potential of this effort.  KELT-1b \citep{siv12} is a $27M_J$ brown dwarf transiting a V=10.7 star.  KELT-1 is the brightest star known to host a transiting brown dwarf.  KELT-2Ab \citep{beatty12} is a transiting hot Jupiter orbiting a V=8.77 star.  KELT-2A, the brighter of two stars in a visual binary, is the ninth-brightest star known to harbor a transiting planet and the third-brightest discovered to date by a ground-based transit survey.   In this paper we describe the discovery and characterization of a hot Jupiter transiting the bright V=9.8 star TYC 2996-683-1.

\section{Discovery and Follow-up Observations\label{sec:obs}}

The KELT survey has an established process for reducing KELT survey data, extracting light curves, identifying potential transiting planets, and performing follow-up observations.  We provide a brief summary of the KELT reductions in \S \ref{sec:keltobs}; for more details, see \S2 of \citet{siv12}.

\subsection{KELT Observations and Photometry\label{sec:keltobs}}

KELT-3 is in KELT-North survey field 06, which is centered on ($\alpha$=09h:46m:48s, $\delta$=+31d:44m:37s; J2000).  We monitored field 06 from October 27, 2006, to April 1, 2011, collecting a total of 6,619 observations. We reduced the raw survey data using a custom implementation of the ISIS image subtraction package \citep{al98, al00}, combined with point-spread fitting photometry using DAOPHOT \citep{stet87}. Using proper motions from the Tycho-2 catalog \citep{hog00} and a reduced proper motion cut \citep{gm03} based on \citet{cc07b}, we selected likely dwarf and subgiant stars within the field for further post-processing and analysis. We applied the trend filtering algorithm \citep[TFA;][]{k05} to each remaining light curve to remove systematic noise, followed by a search for transit signals using the box-fitting least squares algorithm \citep[BLS;][]{k02}. For both TFA and BLS we used the versions found in the VARTOOLS package \citep{hart08}.

\begin{figure}[ht]
\begin{center}
\includegraphics[scale=0.35]{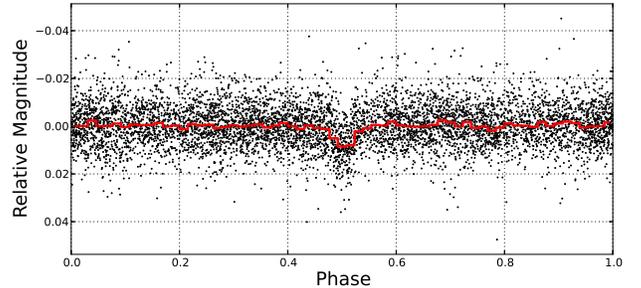}
\caption{Discovery light curve of KELT-3b from the KELT-North telescope.  The light curve contains 6,619 observations spanning 4.4 years, phase-folded to the orbital period of 2.70339 days.  The red line represents the same data binned at 1-hour intervals in phase. \label{fig:kelt_lc}}
\end{center}
\end{figure}

One of the candidates from field 06 was star BD+41 2024 / TYC 2996-683-1 / 2MASS J09543439+4023170, located at ($\alpha$=09h:54m:34.388s, $\delta$=+40d:23m:16.98s; J2000).  The star (henceforth KELT-3) has Tycho magnitudes $B_T = 10.397 \pm 0.032$ and $V_T = 9.873 \pm 0.029$ \citep{hog00} (Johnson magnitudes $B = 10.27$ and $V = 9.82$), and passed our initial selection cuts.  The discovery light curve of KELT-3 is shown in Figure \ref{fig:kelt_lc}.  We observed a transit-like feature at a period of 2.70339 days, with a depth of about 10 mmag.

KELT-3 has a faint ($r = 13.3$) nearby stellar neighbor about 3.7 arcseconds to the northeast, SDSS7 J095434.58+402319.6 \citep[SDSS-DR7;][]{sdss09}.  The presence of this object (henceforth SDSS7J095434) complicates some of our analysis, which we address in \S \ref{sec:exofast}.

\subsection{Radial-Velocity Observations\label{sec:RV}}

After KELT-3 was selected as a candidate, we conducted radial-velocity (RV) observations to identify possible false-positive signatures and to determine the RV orbit.  We obtained data using the Tillinghast Reflector Echelle Spectrograph\footnote{http://tdc-www.harvard.edu/instruments/tres/} \citep[TRES;][]{tres}, on the 1.5m Tillinghast Reflector at the Fred L. Whipple Observatory (FLWO) at Mt. Hopkins, AZ.  We observed KELT-3 sixteen times with TRES over two months, from UT February 26, 2012, to UT May 2, 2012.  The spectra had a resolving power of R=44,000, and were extracted following the procedures described by \citet{bu10}.

We also observed KELT-3 with the FIbre-fed Echelle Spectrograph (FIES) on the 2.5\,m Nordic Optical Telescope (NOT) in La Palma, Spain \citep{dj2010}. We acquired 5 FIES spectra between 13 and 17 March 2012 with the high--resolution fiber ($1\farcs3$ projected diameter) with resolving power of R $\approx 67,000$, and wavelength coverage of $\sim$3600-7400\,\AA. We used the wavelength range from approximately 4000-6100\,\AA\, to determine the radial velocities. The exposure times ranged from 7 to 15 minutes, yielding a SNR from 31 to 52 pixel$^{-1}$ (SNR of 49 to 83 per resolution element) in the wavelength region containing the Mgb triplet.  The procedures used to reduce the FIES spectra and extract the radial velocities are those described in \cite{bu10} and the spectral classification of the FIES spectra is described in \cite{bu12}.

We also obtained follow-up RV measurements from the 2.1m telescope at Kitt Peak National Observatory using the R=30,000 Direct Echelle Mode of the EXPERT spectrograph \citep{ge10}.  Four EXPERT RV measurements were acquired between 20 and 24 January 2013.  The exposure time for each observation ranged from 25 to 40 minutess, yielding a SNR $\sim80$ per pixel around $5400$~\AA. Spectra were reduced using an IDL pipeline modified from an early version described in \citet{wang}. Frames were trimmed, bias subtracted, flat-field corrected, aperture-traced and extracted. All the frames were combined together to serve as a star template.  Radial velocities were derived via cross-correlation with this template from the wavelength range $4900-6300$~\AA.  

\begin{figure}[!t]
\begin{center}
\includegraphics[scale=0.9,trim=5mm 0mm 0mm 0mm,clip]{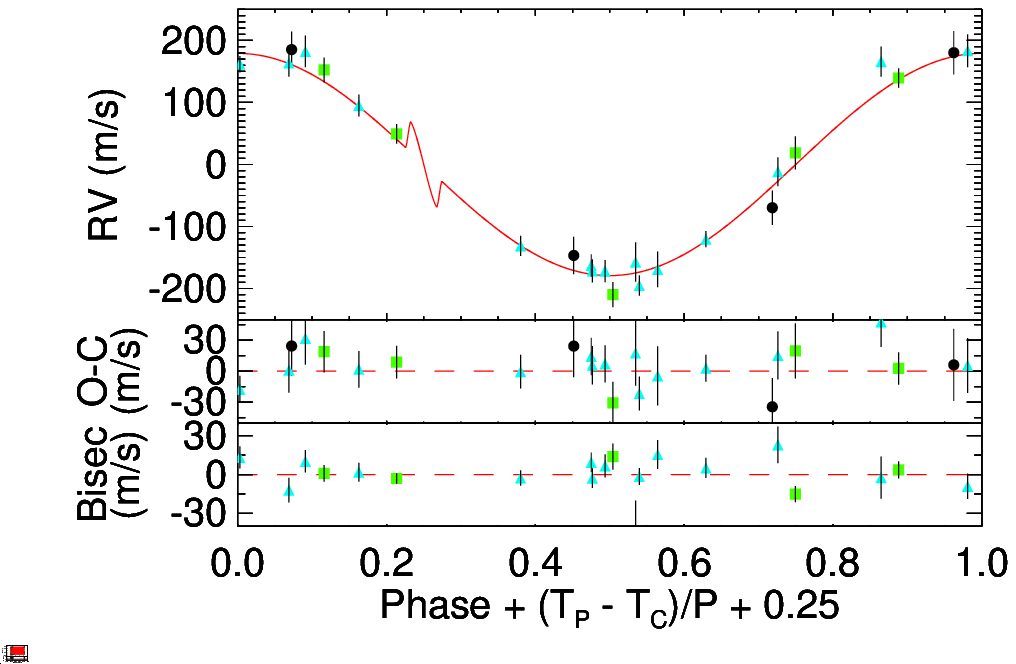}
\caption{Radial velocity measurements of KELT-3. {\it Top panel:} RV observations phased to our best orbital model with eccentricity fixed to zero and with no linear trend, shown in red.  TRES observations are shown as blue triangles, FIES observations are green squares, EXPERT observations are black circles, and the error bars are scaled according to the method described in \S \ref{sec:exofast}.  The RV observation taken during transit is not plotted.  The predicted Rossiter-McLaughlin effect in the model shown incorporates an assumption that $\lambda=0$ (i.e. that the projected spin-obit alignment of the system is 0 degrees).  {\it Middle panel:} Residuals of the RV observations to our circular orbital fit.  {\it Bottom panel:} Bisector span of the TRES and FIES observations as a function of phase.\label{fig:rv}}
\end{center}
\end{figure}

Table \ref{tab:rv} lists all RV data for KELT-3, and Figure \ref{fig:rv} shows the RV data phased to the orbit fit, along with the residuals to the model fit. All RV observations were conducted with fibers with diameters smaller than 2.6 arcseconds, so there is no expected significant contamination from SDSS7J095434, which is about 3.7 arcseconds away. We compute the bisector spans for the TRES and FIES observations.  We find the RMS of the bisector spans to be 13.2 m s$^{-1}$, which is significantly lower than the RV semi-amplitude of 182.0 m s$^{-1}$.  The absence of a trend and low RMS of the bisector spans in phase suggests that the measured RV variations are due to real RV variations in the target star.  Although we do not have bisector spans calculated for the EXPERT data, those constitute only 4/25 RV points, so the bisector results from the other instruments should be sufficient to allay any concerns about the origin of the RV signal.

\begin{center}
\begin{table}
\caption{RV Observations of KELT-3\label{tab:rv}}
\begin{tabular}{lrrl}
\tableline
\multicolumn{1}{c}{BJD} & \multicolumn{1}{c}{RV} & \multicolumn{1}{c}{RV error\tablenotemark{a}} & Source \\
\multicolumn{1}{c}{(TDB)} & \multicolumn{1}{c}{(m s$^{-1}$)} & \multicolumn{1}{c}{(m s$^{-1}$)} & \\
\tableline
2455983.672916  &  -319   &   25   &  TRES \\
2455990.777987  &   -66   &   14   &  TRES \\
2456000.473809  &   -31   &   27   &  FIES \\
2456001.465355  &   103   &   20   &  FIES \\
2456002.514673  &  -259   &   20   &  FIES \\
2456003.552533  &    90   &   16   &  FIES \\
2456004.432251  &     0   &   16   &  FIES \\
2456018.656206  &  -324   &   14   &  TRES \\
2456019.709404  &     4   &   19   &  TRES \\
2456020.710562\tablenotemark{b}  &  -125   &   18   &  TRES \\
2456021.776310  &  -282   &   10   &  TRES \\
2456022.786222  &     0   &   10   &  TRES \\
2456023.806566  &  -293   &   13   &  TRES \\
2456024.741118  &  -173   &   19   &  TRES \\
2456025.668112  &     2   &   18   &  TRES \\
2456026.816586  &  -334   &   14   &  TRES \\
2456029.644415  &  -357   &   13   &  TRES \\
2456033.837821  &    21   &   20   &  TRES \\
2456045.694187  &  -333   &   15   &  TRES \\
2456048.634878  &  -331   &   23   &  TRES \\
2456049.760849  &    22   &   21   &  TRES \\
2456311.939539  &     7   &   29   &  EXPERT \\
2456313.983608  &  -243   &   23   &  EXPERT \\
2456314.940516  &    12   &   24   &  EXPERT \\
2456315.965642  &  -320   &   25   &  EXPERT \\
\tableline
\end{tabular}
\tablenotetext{a}{Unrescaled measurement errors.}
\tablenotetext{b}{This observation occured during transit and was excluded from the EXOFAST analysis in \S \ref{sec:exofast}.}
\end{table}
\end{center}

\subsection{Follow-up Time-Series Photometry} \label{sec:phot}

\begin{figure}[ht]
\begin{center}
\includegraphics[scale=0.85,trim=0mm 0mm 0mm -165mm,clip]{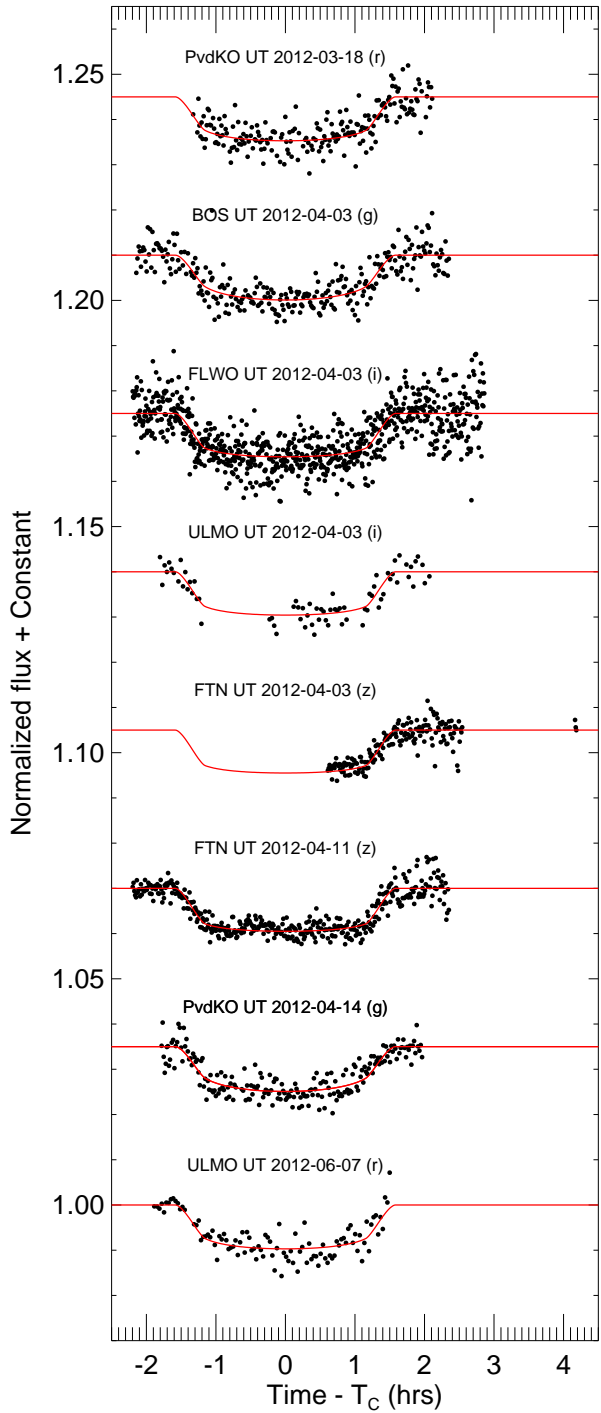}
\caption{Follow-up transit photometry of KELT-3.  The red overplotted lines are the best fit transit model.  The labels are as follows: PvdKO - Peter van de Kamp Observatory (Swarthmore); ULMO - University of Louisville Moore Observatory; FTN - Faulkes Telescope North (LCOGT); BOS - Byrne Observatory at Sedgwick (LCOGT); FLWO - KeplerCam at Fred L. Whipple Observatory.\label{fig:followup_lcs}}
\end{center}
\end{figure}

We acquired follow-up time-series photometry of KELT-3 to check for other types of false positives and to better determine the transit shape.  To schedule followup photometry, we used the Tapir software package\footnote{https://github.com/elnjensen/Tapir, submitted to the Astrophysics Source Code Library}.  We obtained eight partial or full transits in multiple bands between March and June 2012.  For all instruments except FTN, the faint neighbor SDSS7J095434 was included in the photometric aperture, and so the resultant light curves include both KELT-3 and SDSS7J095434.  Figure \ref{fig:followup_lcs} shows all the follow-up light curves assembled.  We find that the $R_P/R_*$ ratio is the same to within the measurement errors in all light curves, which include observations in the $g$, $r$, $i$, and Pan-Starrs-$Z$ filters\footnote{In all references to SDSS filters in this paper, we use the unprimed notation, to denote generic SDSS-like filters, which in practice are often labeled with the primed notation.}, helping to rule out false positives due to blended eclipsing binaries.

We observed two transits of KELT-3 at Swarthmore College's Peter van de Kamp Observatory.  The observatory uses a 0.6m RCOS telescope with an Apogee U16M 4K $\times$ 4K CCD, giving a 26' $\times$ 26' field of view. Using 2 $\times$ 2 binning, it has 0.76 arcseconds pixel$^{-1}$.  On UT 2012-03-17 we obtained a partial transit including egress in $r$.
On UT 2012-04-13 we observed an entire transit in $g$.  

We observed two transits of KELT-3 at Moore Observatory, operated by the University of Louisville.  We used the 0.6m RCOS telescope with an Apogee U16M 4K $\times$ 4K CCD, giving a 26' $\times$ 26' field of view and 0.39 arcseconds pixel$^{-1}$.  The data were calibrated with the AstroImageJ package (Collins, et al., in preparation).  On UT 2012-04-02 we obtained a full transit in $i$, although observing conditions created some interruptions in the data.  On UT 2012-06-07 we obtained a nearly-full transit in $r$.  

We observed two transits with Faulkes Telescope North (FTN), operated by Las Cumbres Observatory Global Telescope (LCOGT).   FTN is a 2.0m telescope with a 4K $\times$ 4K Spectral camera, and we bin the transit observations in 2 $\times$ 2 mode.  On UT 2012-04-02 we obtained a partial transit in Pan-Starrs-$Z$, which includes most of the in-transit phase plus the full egress.  On UT 2012-04-12 we obtained a full transit in Pan-Starrs-$Z$.  Both observations were able to resolve KELT-3 from SDSS7J095434.

We observed a full transit from Byrne Observatory at Sedgwick (BOS), also operated by LCOGT.  BOS is an 0.8m telescope with an 3K $\times$ 2K SBIG STL-6303E camera with 0.572 arcseconds pixel$^{-1}$ and a 14.7' $\times$ 9.8' field of view.  On UT 2012-04-03 we obtained a full transit in $g$, using 2 $\times$ 2 binning.

We observed a full transit in $i$ on UT 2012-04-03 with KeplerCam on the 1.2m telescope at FLWO. KeplerCam has a single 4K $\times$ 4K Fairchild CCD with 0.366 arcseconds pixel$^{-1}$, and a field of view of 23.1' $\times$ 23.1'.  The data were reduced using procedures outlined in \citet{carter11}, which uses standard IDL routines.

\begin{figure}
\begin{center}
\includegraphics[scale=0.85]{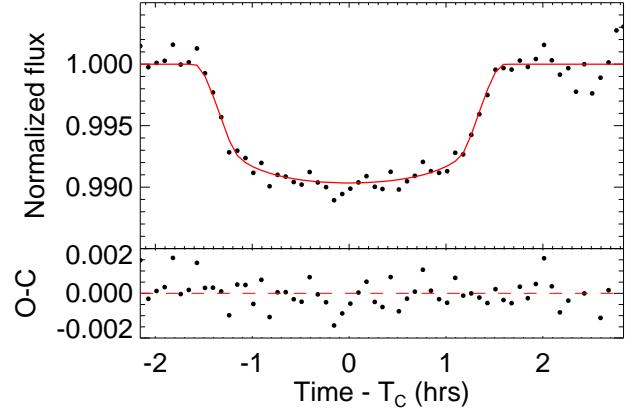}
\caption{{\it Top panel:} All follow-up light curves from Figure \ref{fig:followup_lcs}, combined and binned in 5 minute intervals.  This light curve is not used for analysis, but is shown in order to illustrate the statistical power and behavior of the combined light curve dataset.  The red curve shows the six transit models for each of the individual fits combined and binned in 5 minute intervals the same way as the data, with the model points connected. {\it Bottom panel:} The residuals of the binned light curve from the binned model in the top panel.\label{fig:bestlc}}
\end{center}
\end{figure}

\subsection{Single-epoch Multiband Photometry of SDSS7J095434} \label{sec:nphot}

In an effort to better characterize the relative colors and magnitudes of KELT-3 and SDSS7J095434, we observed the stars with FTN on UT 2012-04-02 in good conditions in the $g$, $r$, $i$ and Pan-Starrs-$Z$ filters.  We find that the neighbor has $g = 14.05 \pm 0.23$, $r = 13.35 \pm 0.17$, $i = 12.85 \pm 0.17$, and $PS-Z = 12.48 \pm 0.19$.    


\subsection{Adaptive Optics Observations\label{sec:ao}}

\begin{figure}
\begin{center}
\fbox{\includegraphics[scale=0.75]{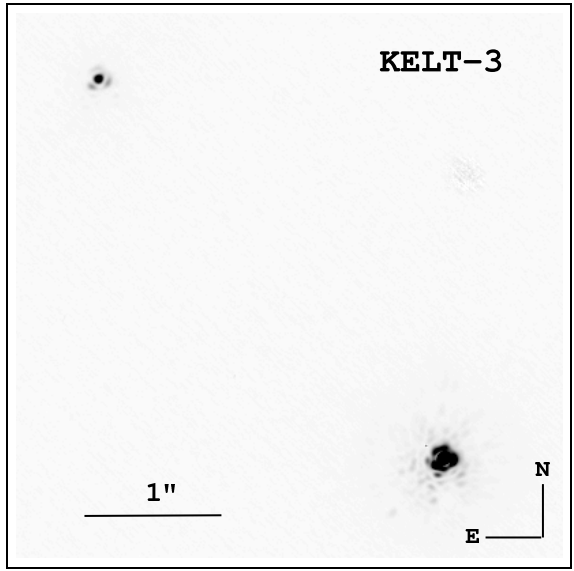}}
\caption{Keck adaptive optics image of KELT-3 taken with NIRC2 in the K' filter.  KELT-3 is in the lower right corner of the panel, while the nearby star SDSS7J095434 is in the upper left, located $3.742 \pm 0.001$ arcseconds to the north east.\label{fig:ao}}
\end{center}
\end{figure}

In order to better assess the nature of SDSS7J095434 and search for any other faint neighbors, we obtained adaptive optics images using NIRC2 (instrument PI: Keith Matthews) at Keck on UT 2012-05-07. Our observations consist of dithered frames taken with the $K$' and $J$ filters. We used the narrow camera setting to provide fine spatial sampling of the stellar point-spread function, and used KELT-3 as its own on-axis natural guide star. The total on-source integration time was 16.3 seconds in each bandpass.  The resulting $K$' image is shown in Figure \ref{fig:ao}.

We find no other faint neighbors in the immediate vicinity of KELT-3.  Specifically, we can exclude additional companions beyond a distance of 0.5 arcseconds from KELT-3 down to a magnitude difference of 6.5 magnitudes at $10$--$\sigma$ confidence.  The magnitude differences between KELT-3 and SDSS7J095434 are $\Delta J$ = $3.001 \pm 0.019$ mag, and $\Delta K$' = $2.436 \pm 0.014$ mag.  The angular separation between the stars is $3724 \pm 1$ mas and the position angle of SDSS7J095434 is $42.00 \pm 0.03$ degrees, measured east of North.

\section{System Parameters and Fits\label{sec:star}}

\subsection{Spectroscopic Analysis\label{sec:spec_params}}

We use both the TRES and FIES spectra to derive the stellar properties of KELT-3.  Using the Spectral Parameter Classification (SPC) procedure \citep{bu12}, we obtained stellar parameters from the average of all 16 TRES spectra, and separately from all 5 FIES spectra.  Since each dataset yielded similar results, we combined the data from all 21 spectra, which yielded from SPC the following results: $T_{\rm eff} = 6308 \pm 50$ K, log(g) = $4.23 \pm 0.10$, [m/H] = $0.04 \pm 0.08$, and $V_{\rm rot} = 10.2 \pm 0.5$ km s$^{-1}$, giving the star an inferred spectral type of F7V.  Based on those results, we are able to calculate the stellar mass and radius from the relations of \citet{torres10}.  In this case, we used [m/H] as a substitute for [Fe/H], but we do not believe that the difference should affect the results.  We find that $M_* = 1.28 \pm 0.12 M_{\odot}$, and $R_* = 1.44 \pm 0.21 R_{\odot}$.

\subsection{SED Analysis\label{sec:sed}}

\begin{figure}
\begin{center}
\includegraphics[scale=0.4,trim=0mm 0mm 0mm 15mm,clip,angle=90]{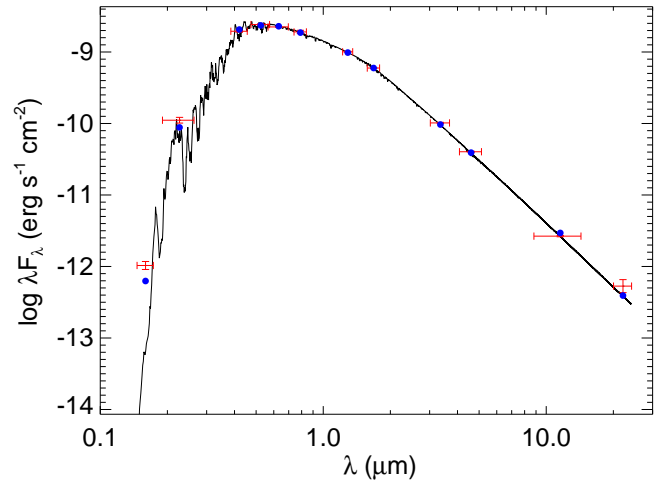}
\caption{Measured and best-fit SED for KELT-3 from UV through NIR.  The red errorbars indicate measurements of the flux of KELT-3 in UV, optical, and NIR passbands listed in Table \ref{tab:starprops}. The vertical errorbars are the $1$--$\sigma$ photometric uncertainties, whereas the horizontal errorbars are the effective widths of the passbands. The solid curve is the best-fit theoretical SED from the NextGen models of \citet{hau99}, assuming stellar parameters $T_{\rm eff}$, log(g) and [Fe/H] fixed at the values in Table \ref{tab:params} from the circular fit, with $A_V$ and $d$ allowed to vary. The blue dots are the predicted passband-integrated fluxes of the best-fit theoretical SED corresponding to our observed photometric bands.\label{fig:sed}}
\end{center}
\end{figure}

We construct an empirical spectral energy distribution (SED) of KELT-3 using the FUV and NUV bandpasses from GALEX \citep{galex05}, the $B_T$ and $V_T$ colors from the Tycho-2 catalog \citep{hog00}, near-infrared (NIR) fluxes in the $J$ and $H$ passbands from the 2MASS Point Source Catalog \citep{2mass03,2mass06}, and near- and mid-IR fluxes in the four WISE passbands \citep{wise10}, to derive the SED shown in Figure \ref{fig:sed}. We fit this SED to NextGen models from \citet{hau99} by fixing the values of $T_{\rm eff}$, log(g) and [Fe/H] inferred from the global fit to the light curve and RV data as described in \S \ref{sec:exofast} and listed in Table \ref{tab:params} for the circular orbit, and then finding the values of the visual extinction $A_V$ and distance $d$ that minimize $\chi^{2}$.  We find $A_V$ = $0.02 \pm 0.02$ and $d$ = $178 \pm 16$ pc.  We note that the quoted statistical uncertainties on $A_V$ and $d$ are likely to be underestimated because we have not accounted for the uncertainties in values of $T_{\rm eff}$, log(g) and [Fe/H] used to derive the model SED. Furthermore, it is likely that alternate model atmospheres would predict somewhat different SEDs and thus values of the extinction and distance.

We also evaluate the motion of KELT-3 through the Galaxy to place it among standard stellar populations.  The TRES RV observations show that it has a bulk radial velocity of $+27.9 \pm 0.2$ km s$^{-1}$.  Combining that with the distance and proper motion information from the NOMAD catalog \citep{nomad04}, we find that KELT-3 has 3-space motion of $U,V,W$ (where positive $U$ is in the direction of the Galactic center) of \,$-25.6 \pm 1.6$, $-11.0 \pm 2.2$, $14.6 \pm 1.3$, all in units of km s$^{-1}$, making it a thin disk star.

\begin{figure}
\begin{center}
\includegraphics[scale=0.5,trim=5mm 0mm 0mm 0mm,clip,angle=0]{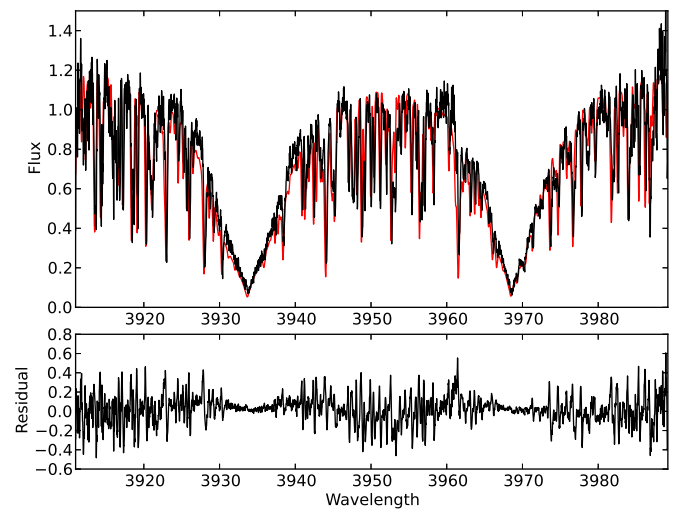}
\caption{No evidence is seen for chromospheric activity in the spectrum of KELT-3.  {\it Upper Panel}: The black curve shows the combined spectra for KELT-3 from the FIES observations, while the red line shows a synthetic spectrum generated using SME.  The spectral range encompasses the Ca H and Ca K absorbtion lines, and no emission is seen in the core from either spectrum, consistent with no signs of chromospheric activity in KELT-3.  {\it Lower Panel}: Residuals between the observed and synthetic spectra.  The scatter in this plot mostly comes from missing line information in the line list for the synthetic spectrum and not noise in either spectrum. The noise level can be seen in the central parts of the Ca lines.\label{fig:rhk}}
\end{center}
\end{figure}

We checked the signatures of chromospheric activity in KELT-3 from the spectroscopic observations, but we find no visible emission in the cores of the Ca-II H and K absorption lines, suggesting an essentially inactive star.  That analysis is shown in Figure \ref{fig:rhk}.  The observed spectrum is compared with a synthetic spectrum generated using SME \citep{sme1,sme2} based on the stellar parameters from \S \ref{sec:exofast}, model atmospheres from \citet{kuru92}, and the line list from \citet{kup00}.  

\subsection{Evolutionary Analysis\label{sec:evolv}}

\begin{figure}
\begin{center}
\includegraphics[scale=0.4,trim=0mm 0mm 0mm 15mm,clip,angle=90]{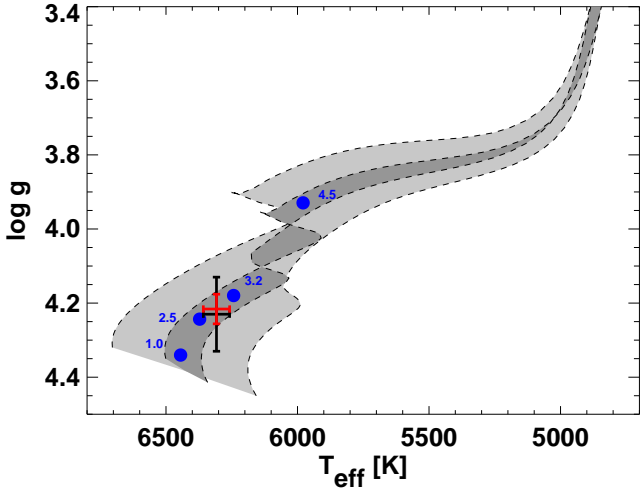}
\caption{Theoretical HR diagram based on Yonsei-Yale stellar evolution models \citep{yy04}. The gray swaths represent
the evolutionary track for the best-fit values of the mass and metallicity of the host star from the circular joint fit described in \S \ref{sec:exofast}, $M_* = 1.282_{-0.060}^{+0.062} M_{\odot}$ and [Fe/H] =$0.048_{-0.081}^{+0.079}$ (dark shaded), and from the spectroscopic constraints alone (light shaded). The tracks for the extreme range of the $1$--$\sigma$ uncertainties on $M_*$ and [Fe/H] from the spectroscopic data only and from the final analysis are shown as dashed lines bracketing the light and dark gray swaths, respectively. The red cross shows the best-fit $T_{\rm eff} = 6304 \pm 49$K and log(g) = $4.204_{-0.029}^{+0.031}$ from the final EXOFAST analysis. The black cross shows the inferred $T_{\rm eff}$ and log(g) from the spectroscopic analysis alone. The blue dots represent the location of the star for various ages in Gyr. The host star is slightly evolved with a probable age of $3.0 \pm 0.02$ Gyr. \label{fig:hrd}}
\end{center}
\end{figure}

In Figure \ref{fig:hrd} we plot the predicted evolutionary track of KELT-3 on a theoretical HR diagram (log(g) vs. $T_{\rm eff}$), from the Yonsei-Yale stellar models \citep{yy04}. Here we have used the stellar mass and metallicity derived from the global circular orbit fit (\S \ref{sec:exofast} and Table \ref{tab:params}).  We also show evolutionary tracks for masses corresponding to the $\pm1$--$\sigma$ extrema in the estimated uncertainty.  We compare our $T_{\rm eff}$ and log(g) values and associated uncertainties to these tracks to  estimate the age of KELT-3. These intersect the evolutionary track around $3.0 \pm 0.2$ Gyr.

To check that the isochrone age is consistent with the other parameters of KELT-3, we calculate the rotation period of the star, using the projected rotational velocity from \S \ref{sec:spec_params} and the stellar radius from the full EXOFAST analysis in \S \ref{sec:exofast}.  Based on that rotation period of $P_{rot}/\sin i_{rot} = 7.11 \pm 0.54$ days, and the colors of the star, we calculate the maximum predicted age (subject to the inclination of the rotation axis to our line of sight) from the models of \citet{barnes07}, which comes to $2.3 \pm 0.7$ Gyr, which is fully consistent with the isochrone age.  We also checked the KELT light curve for periodic variability associated with spot modulation as an independent measure of $P_{rot}$, but we were unable to detect any significant sinusoidal variability beyond the photometric noise.  

\subsection{Characterizing the Faint Neighbor SDSS7J095434\label{sec:nsed}}

\begin{figure}
\begin{center}
\includegraphics[scale=0.4,trim=0mm 0mm 0mm 5mm,clip,angle=90]{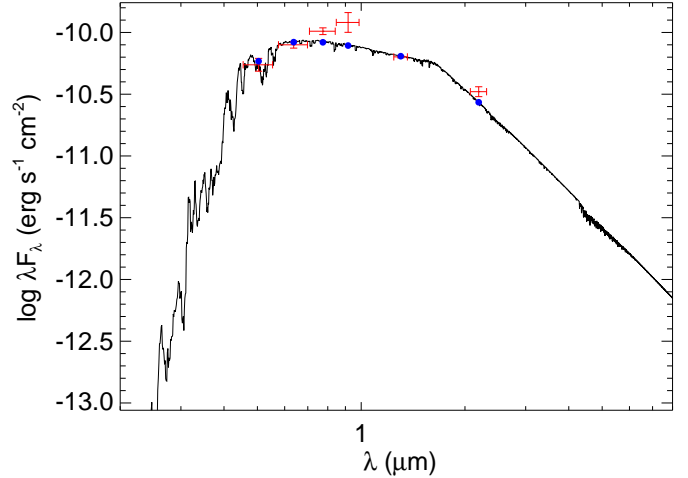}
\caption{SED fit for SDSS7J095434, similar to Figure \ref{fig:sed}.  The red errorbars indicate measurements of the flux of SDSS7J095434 listed in \S \ref{sec:nphot} and \S \ref{sec:ao}. The vertical errorbar indicates the photometric uncertainty, whereas the horizontal errorbar indicates the effective width of the passband. The solid curve is the best-fit theoretical SED from the NextGen models of \citet{hau99}. The blue dots are the predicted passband-integrated fluxes of the best-fit theoretical SED.\label{fig:nsed}}
\end{center}
\end{figure}

Although SDSS7J095434 is identified in the SDSS catalog, its proximity to the much brighter KELT-3 means that the SDSS catalog magnitudes are unreliable, necessitating the new photometric measurements we obtained for this star listed in \S \ref{sec:nphot}.  Based on those measurements and the flux ratios in \S \ref{sec:ao}, we are able to construct an SED for SDSS7J095434 (Figure \ref{fig:nsed}).  We follow the same procedure described in \S \ref{sec:sed}.  Since we do not have independent measures of $T_{\rm eff}$, log(g), and [Fe/H], in this case we let those three parameters vary, along with $A_V$ and $d$, although we limit the maximal value of $A_V$ to 0.04 mag, which is the line-of-sight value based on the dust maps in \citet{dust}.  The colors of SDSS7J095434 suggest it is a K3 star with $T_{\rm eff} = 4800 \pm 400$K, and $A_V = 0.02 \pm 0.02$.  The corresponding distance for such a star with the observed apparent magnitude, assuming it is a dwarf, is $265_{-40}^{+47}$ pc, compared to a distance of $178 \pm 16$ pc for KELT-3.
If SDSS7J095434 is rather a subgiant or giant star, it is located at a much further distance and thus certainly unassociated with KELT-3.
Although the calculated distances to SDSS7J095434 and KELT-3 do not agree to within the quoted $1\sigma$ uncertainties, it is nevertheless possible (and perhaps likely) that they are associated.  Assuming that the uncertainties in the distances are Gaussian distributed, we find that the inferred distances are consistent at the 1.76\% level, assuming SDSS7J095434 is a dwarf star.  Furthermore, as we argued above, it is likely that the uncertainties in the distances are underestimated. 


\subsection{EXOFAST Analysis\label{sec:exofast}}

To determine the final orbital and physical parameters of the KELT-3 system, we use the results from the spectroscopic and SED analyses, the light curves, and the RVs of KELT-3 as inputs to EXOFAST \citep{exo13}, which does a simultaneous Markov Chain Monte Carlo (MCMC) analysis of the entire system, including constraints on the stellar parameters $M_*$ and $R_*$ from the empirical relations in \citet{torres10}. This method is similar to that described in detail in \citet{siv12}, but we note a few differences below.\footnote{In the EXOFAST analysis, which includes the modelling of the filter-specific limb darkening parameters of the transit, we employ the transmission curves defined for the primed SDSS filters rather than the unprimed versions.  We expect any differences due to that discrepancy to be well below the precision of all our observations in this paper and of the limb darkening tables from \citet{cb11}.}

We scale the errors on the RV data such that the probability that the $\chi^2$ was larger than the value we achieved, P($\chi^2$) was 0.5 so as to ensure the resulting parameter uncertainties were roughly accurate. We must do that process separately for the TRES and FIES RV observations, and since there are not enough observations from FIES or EXPERT to perform an independent fit, we fit the TRES data independently, scale their errors so P($\chi^2$) = 0.5, then iteratively scale the FIES errors until the combined P($\chi^2$) = 0.5, and repeat the same process for the EXPERT data. In addition, one of the TRES RVs was taken during the expected time of transit, and we therefore discarded that point so as not to be biased by the Rossiter-McLaughlin effect.

In this analysis, a complication arises because the light from KELT-3 and SDSS7J095434 was blended in all light curves except those from FTN.  As in our analysis of KELT-2Ab \citep{beatty12}, which describes the process in more detail, we iteratively used the primary properties inferred from the full EXOFAST fit, combined with the model SED of the neighbor, to model and subtract the contribution of the flux from the neighbor in all the blended light curves.

The stellar and planetary parameters derived from this procedure are shown in Table \ref{tab:params}.  We see no evidence for a significant slope in the RV data -- leaving the RV slope as a parameter to fit in the EXOFAST analysis yields a slope of $-0.05 \pm 0.36$ m s$^{-1}$ day$^{-1}$.  The long baseline of the combined data sets does not constrain the slope as much as one might naively think. Since separate zero points are fit for each data set, the covariance between these zero points and slope is large, but it is taken into account in our stated uncertainty of the slope, which accounts for a drift in center of mass velocity with time. However, in our final analysis, when we fixed the slope to zero we assume no drift of the center of mass with time, and the entries for gamma in Table \ref{tab:params} are only relative values on an arbitrary scale.  After recomputing the fit with the slope fixed to zero, we find a non-zero eccentricity that is apparently significant at the $\sim 2$--$\sigma$ level, given the formal uncertainty on the eccentricity.   However, as pointed out by \citet{lsbias}, there is a bias for small inferred values of the eccentricity, due to the fact that $e$ is a positive definite quantity.  Therefore, the true significance of a non-zero eccentricity is typically lower than one would infer based on the formal error and assuming a normal distribution.  In this case, the fact that the values of $\ecosw$ and $\esinw$ are both nearly consistent with zero suggests that the inferred non-zero eccentricity is indeed not significant.
To further test the believably of the eccentricity, we performed a prayer bead analysis on the TRES data. We chose 100,000 random links from the eccentricity-fit Markov chain. 
We subtracted the corresponding model from our data, randomly permuted the epochs of residuals using a cyclical boundary, added the residuals back to a model with the same orbital parameters, but with the eccentricity set to zero, and then refit this artificial data set, allowing for eccentricity. Of these fits, 31.5\% preferred an eccentricity of at least the value we measured in the original data set of 0.202, further suggesting this is an artifact of Lucy-Sweeney bias, the observed times, and red noise in residuals rather than a real, non-zero eccentricity. We therefore compute a second EXOFAST fit in which both the orbital eccentricity and the RV slope are fixed to zero, which we prefer, although we quote both fits for completeness.

Using the circular fit with no slope, our final uncertainty scalings are: 1.24 for the TRES RV data, 0.99 for the FIES RV data, and 1.20 for the EXPERT RV data.  The RMS of the RV residuals of the fit to these scaled data are: 20.2 m s$^{-1}$ for TRES, 22.7 m s$^{-1}$ for FIES, and 18.3 m s$^{-1}$ for EXPERT.    We also compute the RMS of the bisectors for the TRES and FIES spectra, obtaining an RMS of 14.3 m s$^{-1}$ for the TRES data, and an RMS of 10.6 m s$^{-1}$ for the FIES data.  As noted previously, we did not measure bisectors for the EXPERT data.

\begin{figure}
\begin{center}
\includegraphics[scale=1.0,trim=10mm 0mm 0mm 5mm,clip,angle=0]{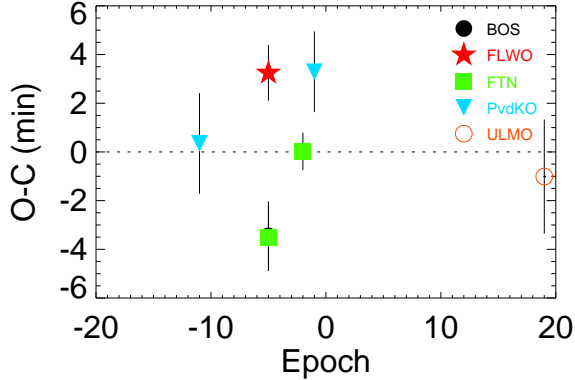}
\caption{The residuals of the transit times from the best-fit ephemeris. The transit times are given in Table \ref{tab:ttvs}.  The BOS observation at Epoch -5 is hidden behind the FTN observation at the same Epoch.\label{fig:ttvs}}
\end{center}
\end{figure}

We investigate the residuals of the circular fit for any signs of transit time variations (TTVs). When we fit the transits shown in Table \ref{tab:ttvs}, the constraints on $T_C$ and $P$ only come from the RV and the prior imposed from the KELT discovery data, not the follow up light curves. Using the transit times to constrain the period during the fit would artificially reduce any observed TTV signal. We fit a straight line to all mid transit times, listed in Table \ref{tab:ttvs} and plotted in Figure \ref{fig:ttvs}, to derive a separate ephemeris from only the transit data: $T_0 = 2456034.29537 \pm 0.00038$, $P = 2.703418 \pm 0.000065$, with a $\chi^2$ of 24.96 and 5 degrees of freedom. While the $\chi^2$ is much larger than one would naively expect, this is largely dominated by the three transits at epoch -5, and in particular the FLWO transit, which is highly discrepant with the other two.  We performed two tests to ensure that the large transit time offsets were not due to problems with the timestamps (e.g., \citep{east10}).  First, we have carefully checked that all quoted times are in BJD$_{TDB}$, re-deriving times from each observer directly from the data in the image headers.  Second, we checked the accuracy of the observatory clocks, and found these to be good to at least 1 second, although these tests were done long after the observations analyzed here were taken.  Therefore, we find no good reason to suspect a problem with the timestamps of the data. The transits shown in Figure \ref{fig:followup_lcs} do not seem dominated by red noise, so it is worrisome that we could find such a large discrepancy (nearly 7 minutes or $4$--$\sigma$) in the transit times at the same epoch. Regardless of the cause, it must be terrestrial in nature, so we conclude there is no convincing evidence for TTVs.

The final system parameters found for KELT-3, based on the circular, no-slope fit, are $M_* = 1.278_{-0.061}^{+0.063} M_{\odot}$, $R_* = 1.472_{-0.067}^{+0.065} R_{\odot}$, $T_{\rm eff} = 6306_{-49}^{+50}$ K, log(g) = $4.209_{-0.031}^{+0.033}$, and [Fe/H] = $0.044_{-0.082}^{+0.080}$
The planet is an inflated hot Jupiter with $M_P = 1.477_{-0.067}^{+0.066} M_J$, and $R_P = 1.345 \pm 0.072 R_J$,
It is strongly irradiated, with an equilibrium temperature of $T_{\rm eq} = 1816_{-39}^{+37}$ K and an incident flux of $2.47_{-0.20}^{+0.21} \times 10^9$ erg s$^{-1}$ cm$^{-2}$.  The radius is about 25\% larger than predicted from the models of \citet{baraffe} for an irradiated planet with a mass of 1 to 2 $M_J$ at 3 Gyr and a small amount of metals.  

\section{Backtracking the Evolution of KELT-3\label{sec:kalo}}

KELT-3b is somewhat inflated, with a density of $0.75_{-0.10}^{+0.12}$ g cm$^{-3}$.  In an investigation of transiting giant exoplanets, \citet{ds11} found that exoplanets that are insolated beyond a certain threshold ($2 \times 10^8$ erg s$^{-1}$ cm$^{-2}$) have radii that are inflated compared to those planets with lower levels of insolation.  KELT-3b falls well above that threshold, and follows the insolation-inflation trend displayed in Figure 1 of \citet{ds11}.
It is worth investigating, however, whether that relationship has always held true.  That is, has KELT-3b always been as insolated as it is now?  If it turns out that KELT-3b only recently began receiving enhanced irradiation, this could provide an empirical probe of the timescale of inflation mechinisms (see \citet{assef09} and \citet{sm12}).

In order to answer that question, we simulate the reverse-evolution of the star-planet system, using the measured parameters listed in this paper as the present boundary conditions.  This analysis is not intended to examine circularization  of the planet, tidal locking to the star, or any type of planet-planet or planet-disk migration.  Rather, it is a way to investigate the change in insolation of the planet over time due to the changing luminosity of the star and changing star-planet separation.

\begin{figure}
\begin{center}
\includegraphics[scale=0.5,trim=0mm 0mm 0mm 0mm,clip,angle=0]{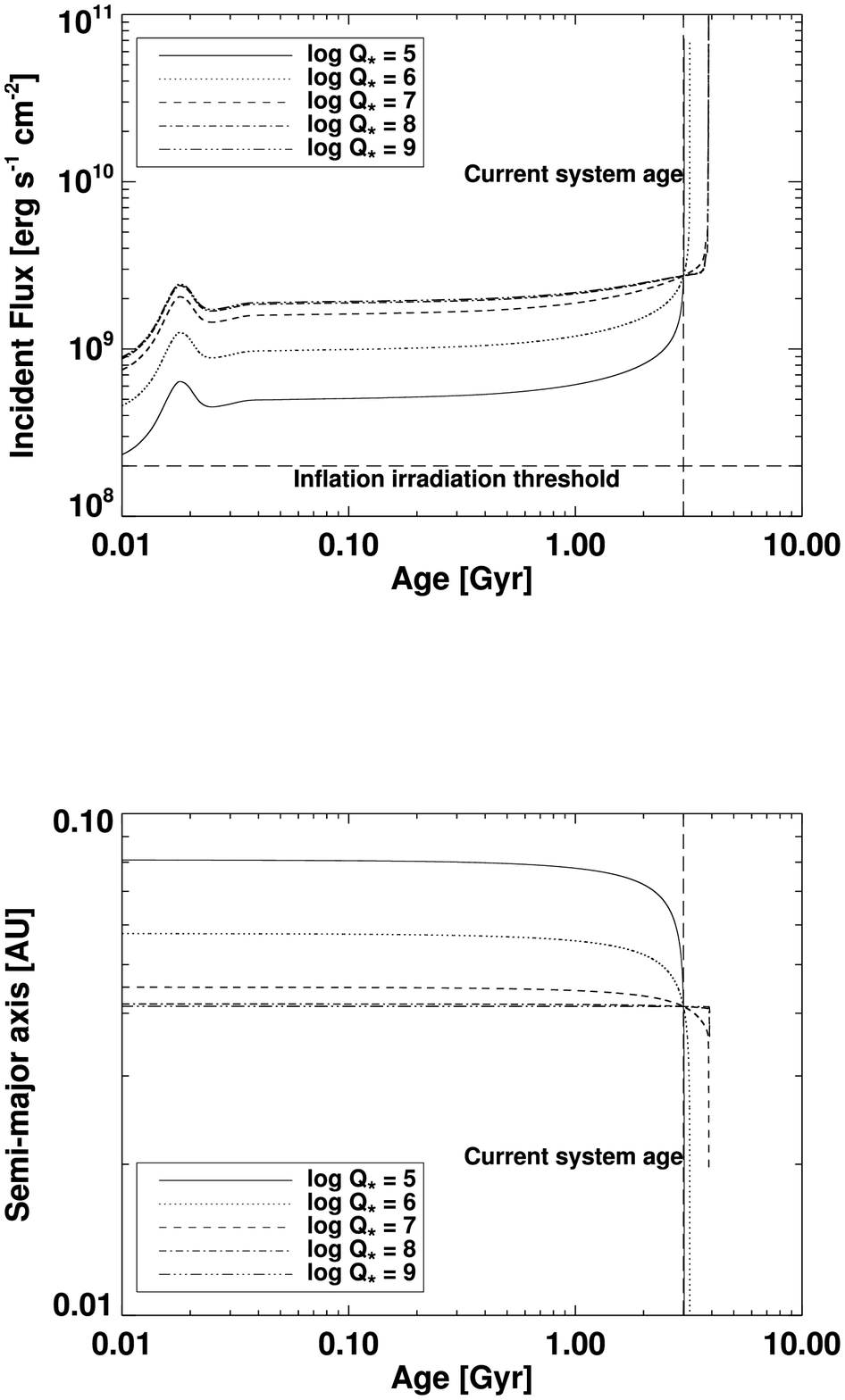}
\caption{Change in incident flux ({\it top}) and semi-major axis ({\it bottom}) for KELT-3b, with different test values for $Q_{\star}$ for KELT-3.  In all cases the planet has always received more than enough flux from its host to keep the planet irradiated beyond the insolation threshold of $2 \times 10^8$ erg s$^{-1}$ cm$^{-2}$ identified by \citet{ds11}.\label{fig:evolv}}
\end{center}
\end{figure}

We include the evolution of the star, assumed to follow the YREC $1.3 M_{\odot}$ model with solar metallicity \citep{siess00}.  We assume that the stellar rotation was influenced only by tidal torques due to the planet, with no magnetic wind and treating the star like a solid body.  We also assume a circular orbit throughout the full analysis.  The results of our simulations are shown in Figure \ref{fig:evolv}.  We tested a range of values for the tidal quality factor of the star $Q_{\star}$, from log($Q_{\star}$) = 5 to log($Q_{\star}$) = 9.  We find that although for certain values of $Q_{\star}$ the planet has moved substantially closer to its host during the past Gyr, in all cases the planet has always received more than enough flux from its host to keep the planet irradiated beyond the insolation threshold identified by \citet{ds11}.  The rapid changes in semimajor axis and incident flux moving toward the future are mainly due to certain $Q_{\star}$ values being unphysically low and the logarithmic layout of the plots in age, and should not be taken as a sign that we are seeing the system at an especially unique moment of evolution.

\section{Discussion\label{sec:discuss}}

\begin{figure}
\begin{center}
\includegraphics[scale=0.45,angle=0]{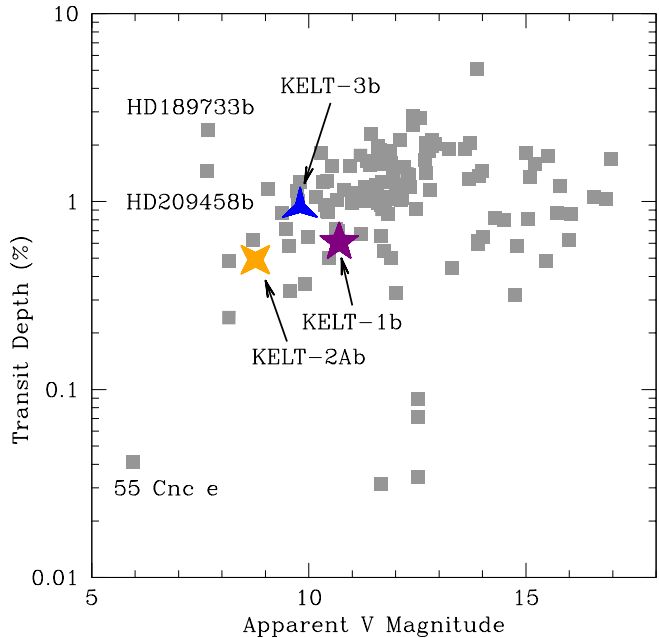}
\caption{Transit depth as a function of the apparent V magnitude of the host star for a sample of transiting systems. KELT-3b is shown as the blue 3-pointed star. All else being equal, objects in the top left provide the best targets for follow-up.   The other discoveries from the KELT survey, KELT-1b and KELT-2Ab, are also shown.\label{fig:v_depth}}
\end{center}
\end{figure}

KELT-3b is a typical hot Jupiter.  Its host star is among the 20 brightest transiting planet hosts, and so the system provides an opportunity for detailed follow-up observations and analysis.  Figure \ref{fig:v_depth} shows the location of KELT-3 in a plot of host star brightness versus transit depth relative to other bright transiting planets hosts, demonstrating the potential value of this system for follow-up studies of its atmosphere.

Aside from its follow-up value as an especially bright transiting planet host, KELT-3 itself is nearly identical to the transiting planet host star HAT-P-2 \citep{hatp2}.  Both stars have the same masses, radii, effective temperatures, and surface gravities to within the measurement errors for both stars, and metallicities that differ by only 0.1 dex.  The only stellar parameter that differs markedly is the rotational velocity, with HAT-P-2 rotating twice as fast as KELT-3 (20.8 km s$^{-1}$ versus 10.2 km s$^{-1}$).  The availabity of such similar planet hosts can be used to compare different planet formation and evolution theories by comparing how closely exoplanet properties track the properties of their host stars.  This particular case offers an especially interesting comparison, since KELT-3b is a relatively typical hot Jupiter, while HAT-P-2b is quite an odd planet companion, with an exceptionally high mass (about $9 M_J$) and eccentricity (0.52).  

\acknowledgments

We would like to thank additional collaborators on the KELT project, including Bruce Gary, Joao Gregorio, Roberto Zambelli, and Kim McLeod.  We would also like to thank Leslie Hebb, Robert Emrich, and Martin Paegert for helpful discussions.
Early work on KELT-North was supported by NASA Grant NNG04GO70G.
J.A.P. and K.G.S. acknowledge support from the Vanderbilt Office of the Provost through the Vanderbilt Initiative in Data-intensive Astrophysics.  
E.L.N.J. gratefully acknowledges the support of the National Science Foundation's PREST program, which helped to establish the Peter van de Kamp Observatory through grant AST-0721386.
K.G.S. acknowledges the support of the National Science Foundation through PAARE Grant AST-0849736 and AAG Grant AST-1009810.
K.A.C. was supported by a NASA Kentucky Space Grant Consortium Graduate Fellowship.  
Work by B.S.G., J.D.E., and T.G.B.\ was partially supported by NSF CAREER Grant AST-1056524.  
The TRES and KeplerCam observations were obtained with partial support from the Kepler Mission through NASA Cooperative Agreement NNX11AB99A with the Smithsonian Astrophysical Observatory (PI: D.W.L.).
The Byrne Observatory at Sedgwick (BOS) is operated by the Las Cumbres Observatory Global Telescope Network and is located at the Sedgwick Reserve, a part of the University of California Natural Reserve System.  
EXPERT construction and follow-up observations were funded by NSF with grant NSF AST-0705139, NASA with grants NNX07AP14G (Origins), UCF-UF SRI program, and the University of Florida. 
This work has made use of NASA's Astrophysics Data System, the Exoplanet Orbit Database at exoplanets.org, the Extrasolar Planet Encyclopedia at exoplanet.eu \citep{epe11}, and the SIMBAD database operated at CDS, Strasbourg, France.

\begin{deluxetable}{llclc}
\tablecolumns{5}
\tablewidth{0pt}
\tabletypesize{\scriptsize}
\tablecaption{Stellar Properties of KELT-3\label{tab:starprops}}
\tablehead{\colhead{~~~Parameter} & \colhead{Description (Units)} & \colhead{Value}  & \colhead{Source} & \colhead{Reference}  }
\startdata
Names              &                                      &    BD+41 2024                 &            &     \\
                   &                                      &    TYC 2996-683-1             &            &     \\
                   &                                      &    2MASS J09543439+4023170    &            &     \\
                   &                                      &    GSC 02996-00683            &            &     \\
                   &                                      &    SAO 43097                  &            &     \\
$\alpha_{J2000}$   &                                      &    09 54 34.391               &  Tycho-2   &   1  \\
$\delta_{J2000}$   &                                      &   +40 23 16.98                &  Tycho-2   &   1  \\
FUV$_{GALEX}$      &                                      &   $17.95 \pm 0.14$            &  GALEX     &   2  \\
NUV$_{GALEX}$      &                                      &   $14.09 \pm 0.01$            &  GALEX     &   2  \\
$B_T$              &                                      &   $10.397 \pm 0.032$          &  Tycho-2   &   1  \\
$V_T$              &                                      &   $9.873 \pm 0.029$           &  Tycho-2   &   1  \\
$r_{SDSS}$         &                                      &   $9.728 \pm 0.015$           &  Carlsberg &   3  \\
$I_{C}$            &                                      &   $9.263 \pm 0.057$           &  TASS      &   4  \\
$J$                &                                      &   $8.963 \pm 0.019$           &  2MASS     &   5  \\
$H$                &                                      &   $8.728 \pm 0.019$           &  2MASS     &   5  \\
WISE1              &                                      &   $11.26  \pm  0.022$         &  WISE      &   6  \\
WISE2              &                                      &   $11.923 \pm  0.019$         &  WISE      &   6  \\     
WISE3              &                                      &   $13.874 \pm  0.023$         &  WISE      &   6  \\  
WISE4              &                                      &   $14.918 \pm  0.227$         &  WISE      &   6  \\  
$\mu_{\alpha}$     & Proper Motion in RA (mas yr$^{-1}$)  &   $-28.9  \pm  0.6$           &  NOMAD     &   7  \\  
$\mu_{\delta}$     & Proper Motion in Dec (mas yr$^{-1}$) &   $-25.1  \pm  0.7$           &  NOMAD     &   7  \\  
$U$\tablenotemark{a}                & ${\rm km~s^{-1}}$                    &   $-25.6 \pm 1.6$             & This paper &     \\  
$V$                & ${\rm km~s^{-1}}$                    &   $-11.0 \pm 2.2$             & This paper &     \\  
$W$                & ${\rm km~s^{-1}}$                    &   $14.6 \pm 1.3$               & This paper &     \\  
$d$                & Distance (pc)                        &   $178 \pm 16$                & This paper &     \\  
                   & Age (Gyr)                            &   $3.0 \pm 0.2$               & This paper &     \\  
$A_V$              & Visual Extinction                    &   $0.02 \pm 0.02$             & This paper &     \\  
\enddata
\tablenotetext{a}{Positive $U$ is in the direction of the Galactic Center.}
\tablerefs{1=\citet{hog00}; 2=\citet{galex05}; 3=\citet{carlsberg06}; 4=\citet{tass2000}; 5=\citet{2mass03,2mass06};
6=\citet{wise10, wise12}; 7=\citet{nomad04}}
\end{deluxetable}

\begin{deluxetable}{llcc}
\tablecaption{Median Values and 68\% Confidence Intervals for the Physical and Orbital Parameters of the KELT-3 System\label{tab:params}}
\tablehead{\colhead{~~~Parameter} & \colhead{Units} & \colhead{Value; $e \neq 0$} & \colhead{Value; $e \equiv 0$, adopted}}
\startdata
\sidehead{Stellar Parameters:}
                               ~~~$M_{*}$\dotfill &Mass (\msun)\dotfill & $1.277_{-0.062}^{+0.063}$  & $1.278_{-0.061}^{+0.063}$\\
                             ~~~$R_{*}$\dotfill &Radius (\rsun)\dotfill & $1.464_{-0.077}^{+0.076}$  & $1.472_{-0.067}^{+0.065}$\\
                         ~~~$L_{*}$\dotfill &Luminosity (\lsun)\dotfill & $3.04_{-0.33}^{+0.35}$     & $3.08_{-0.30}^{+0.31}$\\   
                             ~~~$\rho_*$\dotfill &Density (cgs)\dotfill & $0.574_{-0.070}^{+0.086}$  & $0.565_{-0.059}^{+0.071}$\\
                  ~~~$\log{g_*}$\dotfill &Surface gravity (cgs)\dotfill & $4.213_{-0.036}^{+0.039}$  & $4.209_{-0.031}^{+0.033}$\\
                  ~~~$\teff$\dotfill &Effective temperature (K)\dotfill & $6304\pm50$                & $6306_{-49}^{+50}$\\       
                                 ~~~$\feh$\dotfill &Metallicity\dotfill & $0.046\pm0.080$            & $0.044_{-0.082}^{+0.080}$\\
\sidehead{Planetary Parameters:}
                                   ~~~$e$\dotfill &Eccentricity\dotfill & $0.202_{-0.089}^{+0.079}$     & $\equiv 0$\\
        ~~~$\omega_*$\dotfill &Argument of periastron (degrees)\dotfill & $-158_{-58}^{+48}$            & $\equiv 90$\\
                                  ~~~$P$\dotfill &Period (days)\dotfill & $2.7033902\pm0.0000099$       & $2.7033904_{-0.0000100}^{+0.0000099}$\\            
                           ~~~$a$\dotfill &Semi-major axis (AU)\dotfill & $0.04120\pm0.00067$           & $0.04122_{-0.00067}^{+0.00066}$\\                  
                                 ~~~$M_{P}$\dotfill &Mass (\mj)\dotfill & $1.444_{-0.068}^{+0.069}$     & $1.477_{-0.064}^{+0.066}$\\                        
                               ~~~$R_{P}$\dotfill &Radius (\rj)\dotfill & $1.340\pm0.080$               & $1.345\pm0.072$\\                                  
                           ~~~$\rho_{P}$\dotfill &Density (cgs)\dotfill & $0.75_{-0.11}^{+0.14}$        & $0.75_{-0.10}^{+0.12}$\\                           
                      ~~~$\log{g_{P}}$\dotfill &Surface gravity\dotfill & $3.300_{-0.045}^{+0.047}$     & $3.306_{-0.041}^{+0.044}$\\                        
               ~~~$T_{eq}$\dotfill &Equilibrium temperature (K)\dotfill & $1811_{-44}^{+43}$            & $1816_{-39}^{+37}$\\                               
                           ~~~$\Theta$\dotfill &Safronov number\dotfill & $0.0696_{-0.0043}^{+0.0046}$  & $0.0708_{-0.0040}^{+0.0045}$\\                     
                   ~~~$\fave$\dotfill &Incident flux (\fluxcgs)\dotfill & $2.35\pm0.25$                 & $2.47_{-0.20}^{+0.21}$\\                           
\sidehead{RV Parameters:}
       ~~~$T_C$\dotfill &Time of inferior conjunction (\bjdtdb)\dotfill & $2456023.482_{-0.031}^{+0.039}$              & $2456023.459\pm0.017$\\
               ~~~$T_{P}$\dotfill &Time of periastron (\bjdtdb)\dotfill & $2456024.12_{-0.42}^{+0.40}$                 & --- \\
                        ~~~$K$\dotfill &RV semi-amplitude (m/s)\dotfill & $182.3\pm5.2$                                & $182.0\pm5.2$\\                           
                    ~~~$M_P\sin{i}$\dotfill &Minimum mass (\mj)\dotfill & $1.437_{-0.067}^{+0.068}$                    & $1.470_{-0.064}^{+0.065}$\\               
                           ~~~$M_{P}/M_{*}$\dotfill &Mass ratio\dotfill & $0.001081\pm0.000040$                        & $0.001104\pm0.000036$\\                   
                       ~~~$u$\dotfill &RM linear limb darkening\dotfill & $0.6017_{-0.0054}^{+0.0056}$                 & $0.6014_{-0.0054}^{+0.0056}$\\            
                              ~~~$\gamma_{EXPERT}$\dotfill &m/s\dotfill & $-168_{-20}^{+19}$                           & $-170\pm15$\\                             
                                ~~~$\gamma_{FIES}$\dotfill &m/s\dotfill & $-47.5_{-8.3}^{+8.1}$                        & $-47.1_{-8.3}^{+8.4}$\\                   
                                ~~~$\gamma_{TRES}$\dotfill &m/s\dotfill & $-155.7_{-5.7}^{+5.8}$                       & $-158.7\pm4.9$\\                          
                                         ~~~$\ecosw$\dotfill & \dotfill & $-0.145_{-0.100}^{+0.15}$                    & --- \\
                                         ~~~$\esinw$\dotfill & \dotfill & $-0.06_{-0.11}^{+0.13}$                      & --- \\
                     ~~~$f(m1,m2)$\dotfill &Mass function (\mj)\dotfill & $0.00000166_{-0.00000016}^{+0.00000017}$     & $0.00000177_{-0.00000015}^{+0.00000016}$\\
\sidehead{Primary Transit Parameters:}
~~~$R_{P}/R_{*}$\dotfill &Radius of the planet in stellar radii\dotfill & $0.0941\pm0.0011$                            & $0.0939\pm0.0011$\\               
           ~~~$a/R_*$\dotfill &Semi-major axis in stellar radii\dotfill & $6.05_{-0.26}^{+0.29}$                       & $6.02_{-0.22}^{+0.24}$\\          
                          ~~~$i$\dotfill &Inclination (degrees)\dotfill & $84.25_{-0.64}^{+0.67}$                      & $84.23_{-0.58}^{+0.65}$\\         
                               ~~~$b$\dotfill &Impact parameter\dotfill & $0.612\pm0.072$                              & $0.606_{-0.046}^{+0.037}$\\       
                             ~~~$\delta$\dotfill &Transit depth\dotfill & $0.00885\pm0.00021$                          & $0.00882\pm0.00021$\\             
                    ~~~$T_{FWHM}$\dotfill &FWHM duration (days)\dotfill & $0.1159_{-0.0051}^{+0.0040}$                 & $0.11414_{-0.00081}^{+0.00080}$\\ 
              ~~~$\tau$\dotfill &Ingress/egress duration (days)\dotfill & $0.0178_{-0.0029}^{+0.0037}$                 & $0.0172\pm0.0015$\\               
                     ~~~$T_{14}$\dotfill &Total duration (days)\dotfill & $0.1346_{-0.0086}^{+0.0062}$                 & $0.1313\pm0.0016$\\               
   ~~~$P_{T}$\dotfill &A priori non-grazing transit probability\dotfill & $0.147_{-0.019}^{+0.024}$                    & $0.1504_{-0.0057}^{+0.0055}$\\    
              ~~~$P_{T,G}$\dotfill &A priori transit probablity\dotfill & $0.178_{-0.023}^{+0.029}$                    & $0.1816_{-0.0072}^{+0.0069}$\\    

                ~~~$u_{1Sloang}$\dotfill &Linear Limb-darkening\dotfill & $0.4675_{-0.0089}^{+0.0097}$                 & $0.4672_{-0.0089}^{+0.0098}$\\
             ~~~$u_{2Sloang}$\dotfill &Quadratic Limb-darkening\dotfill & $0.2748_{-0.0050}^{+0.0044}$                 & $0.2749_{-0.0050}^{+0.0044}$\\
                ~~~$u_{1Sloani}$\dotfill &Linear Limb-darkening\dotfill & $0.2347_{-0.0050}^{+0.0053}$                 & $0.2344_{-0.0051}^{+0.0053}$\\
             ~~~$u_{2Sloani}$\dotfill &Quadratic Limb-darkening\dotfill & $0.3103_{-0.0025}^{+0.0026}$                 & $0.3104_{-0.0025}^{+0.0026}$\\
                ~~~$u_{1Sloanr}$\dotfill &Linear Limb-darkening\dotfill & $0.3083_{-0.0059}^{+0.0064}$                 & $0.3080_{-0.0059}^{+0.0064}$\\
             ~~~$u_{2Sloanr}$\dotfill &Quadratic Limb-darkening\dotfill & $0.3195_{-0.0020}^{+0.0021}$                 & $0.3196_{-0.0020}^{+0.0022}$\\
                ~~~$u_{1Sloanz}$\dotfill &Linear Limb-darkening\dotfill & $0.1835_{-0.0046}^{+0.0048}$                 & $0.1833\pm0.0047$\\           
             ~~~$u_{2Sloanz}$\dotfill &Quadratic Limb-darkening\dotfill & $0.3016_{-0.0025}^{+0.0020}$                 & $0.3016_{-0.0025}^{+0.0021}$\\
\sidehead{Secondary Eclipse Parameters:}
                  ~~~$T_{S}$\dotfill &Time of eclipse (\bjdtdb)\dotfill & $2456024.58_{-0.13}^{+0.23}$                 &  $2456022.107\pm0.017$ \\
                           ~~~$b_{S}$\dotfill &Impact parameter\dotfill & $0.551_{-0.086}^{+0.10}$                     &  --- \\
                  ~~~$T_{S,FWHM}$\dotfill &FWHM duration (days)\dotfill & $0.1107_{-0.0093}^{+0.0064}$                 &  --- \\
            ~~~$\tau_S$\dotfill &Ingress/egress duration (days)\dotfill & $0.0152_{-0.0029}^{+0.0046}$                 &  --- \\
                   ~~~$T_{S,14}$\dotfill &Total duration (days)\dotfill & $0.126_{-0.012}^{+0.011}$                    &  --- \\
   ~~~$P_{S}$\dotfill &A priori non-grazing eclipse probability\dotfill & $0.164_{-0.020}^{+0.019}$                    &  --- \\
             ~~~$P_{S,G}$\dotfill &A priori eclipse probability\dotfill & $0.198_{-0.024}^{+0.023}$                    &  --- \\

\enddata
\end{deluxetable}

\begin{table}[ht]
\caption{Transit Times for KELT-3\label{tab:ttvs}}
\begin{tabular}{crrrrrr}
\tableline
Telescope  & Epoch & T$_{\rm C}$      & Error     & O-C       & (O - C)/Error    \\
and filter &       & (\bjdtdb)        & (seconds) & (seconds) &                  \\
\tableline
PvdKO, $f$ & -11   &  2456004.558023  & 123       &   21.02   &  0.17            \\
BOS, $g$   &  -5   &  2456020.775891  &  84       & -206.86   & -2.44            \\
FLWO, $i$  &  -5   &  2456020.780541  &  68       &  194.90   &  2.85            \\
FTN, $z$   &  -5   &  2456020.775835  &  81       & -211.70   & -2.60            \\
FTN, $z$   &  -2   &  2456028.888551  &  46       &    1.12   &  0.02            \\
PvdKO, $g$ &  -1   &  2456031.594242  &  99       &  197.54   &  1.99            \\
ULMO, $r$  &  19   &  2456085.659606  & 140       &  -60.60   & -0.43            \\
\end{tabular}
\end{table}

\end{document}